\documentclass[11pt,letterpaper]{article}
\usepackage{graphicx}
\usepackage{bm}
\usepackage{amssymb}
\usepackage{color}
\usepackage{amsmath}
\usepackage[numbers]{natbib}
\usepackage{enumerate}
\usepackage{amstext}
\usepackage{footmisc}
\usepackage{latexsym}
\usepackage[usenames,dvipsnames]{xcolor}
\usepackage[colorlinks=true,citecolor=Cerulean,linkcolor=RubineRed,urlcolor=Cerulean]{hyperref}
\usepackage{multibib}
\usepackage{ fixfoot}
\usepackage{fancyhdr}
\newcommand{\pac}[1]{ \left\{ #1 \right\} }
\newcommand{\pap}[1]{\left( #1 \right)}
\newcommand{\pas}[1]{\left[#1 \right]}

\newcommand{\tr}[1]{\mathrm{tr}\left\{ #1 \right\}}

\DeclareFixedFootnote{\rep}{Electronic address: \href{mailto:fj.gomez34@dipc.org}{fj.gomez34@dipc.org}}

\def\ii{{\rm i}}
\def\ee{{\rm e}}
\def\lll{{\it  l}}

\def\L{{\rm L}}
\def\J{{\rm J}}

\def\tr{\rm{Tr}}
\def\da{\dagger}
\newcommand{\beq}{\begin{equation}}
\newcommand{\eeq}{\end{equation}}
\newcommand{\beqa}{\begin{eqnarray}}
\newcommand{\eeqa}{\end{eqnarray}}

\begin{document}
\title{{\bf Enhancing violations of Leggett-Garg inequalities in nonequilibrium correlated many-body systems by interactions and decoherence}}
\author{J. J. Mendoza-Arenas,$^{1}$ F. J. G\'omez-Ruiz,$^{2,1,}$\rep{}\: F. J. Rodr\'iguez,$^1$\\
 \& L. Quiroga$^{1}$}
\date{}
\maketitle
\vspace{-1cm}
\begin{center}
$^{1}${\it Departamento de F{\'i}sica, Universidad de los Andes, A.A. 4976, Bogot\'a D. C., Colombia}\\
$^{2}${\it Donostia International Physics Center,  E-20018 San Sebasti\'an, Spain}
\end{center}
\begin{abstract}
We identify different schemes to enhance the violation of Leggett-Garg inequalities in open many-body systems. Considering a nonequilibrium archetypical setup of quantum transport, we show that particle interactions control the direction and amplitude of maximal violation, and that in the strongly-interacting and strongly-driven regime bulk dephasing enhances the violation. Through an analytical study of a minimal model we unravel the basic ingredients to explain this decoherence-enhanced quantumness, illustrating that such an effect emerges in a wide variety of systems.
\end{abstract}

\section*{Introduction}
For several decades, assessing the existence of genuine quantum behavior~\cite{Sanpera:2018rep} and quantifying coherence in quantum systems~\cite{plenio:2018rmp} have remained as fundamental open problems, critical to applications in computation and information processing. A seminal breakthrough in this direction was made by Leggett and Garg~\cite{leggett,emary}, who considered the conditions that a macroscopic system entirely described by classical physics should satisfy. These are macroscopic realism (a system's property is well defined at every time regardless of whether it is observed or not) and noninvasive measurability (the system is unaffected by measurements on it). When these conditions are met, the system satisfies the so-called Leggett-Garg inequalities (LGIs). The experimental observation of their violation, which serves as witness of quantum coherence, has been sought intensively and reported in various platforms~\cite{palacios,goggin,knee,dressel,zhou2015prl}. \\
\\
In order to establish realistic conditions where LGIs are maximally violated, providing optimal settings for quantum protocols performed in a particular system, its unavoidable coupling to the environment has to be considered. Recent theoretical research on few-qubit systems has determined that Markovian noise degrades quantumness~\cite{emary2013pra,lobejko2015pra,Friedenberger:2017pra,chanda2018pra,ban2018qip}, while non-Markovianity helps restore it~\cite{chen2014scirep,naikoo:2018,datta:2018}. However, many-body interacting systems, where spectacular effects resulting from quantum coherence emerge, remain unexplored within this effort. The non-trivial impact of dissipation in multilevel systems is known to induce unexpected beneficial effects, as entanglement generation and quantum state engineering~\cite{diehl2008nat,verstraete2009nat,lin2013nat,kienzler2015sci}, restoration of hidden quantum phase transitions~\cite{zhang:2017} and environment-assisted transport~\cite{plenio2008njp,mohseni_jcp2008,chin2010njp,sinayskiy_prl2012}, recently observed in several quantum simulators~\cite{viciani2015prl,gorman2018prx,potocnik2018nat,maier2019prl}. Thus such systems constitute attractive candidates for uncovering novel mechanisms of LGI violation enhancement.\\
\\
In the present work we establish such mechanisms in testbed open many-body systems. We show that not only particle interactions, but also bulk dephasing, can increase the violation of LGIs in Markovian systems. For this we assess the quantumness of the transport supported by interacting spin chains when driven out of equilibrium by unequal boundary reservoirs, a configuration considered so far for LGIs on single-particle systems only~\cite{lambert2010prl,castillo2013pra}. Using a minimal model to illustrate the basic ingredients responsible for this phenomenon, we argue that it can emerge under a wide range of conditions.

\section*{Methods}
\subsection*{Nonequilibrium setup} 
\begin{figure}[h!]
\begin{center}
\includegraphics[scale=0.75]{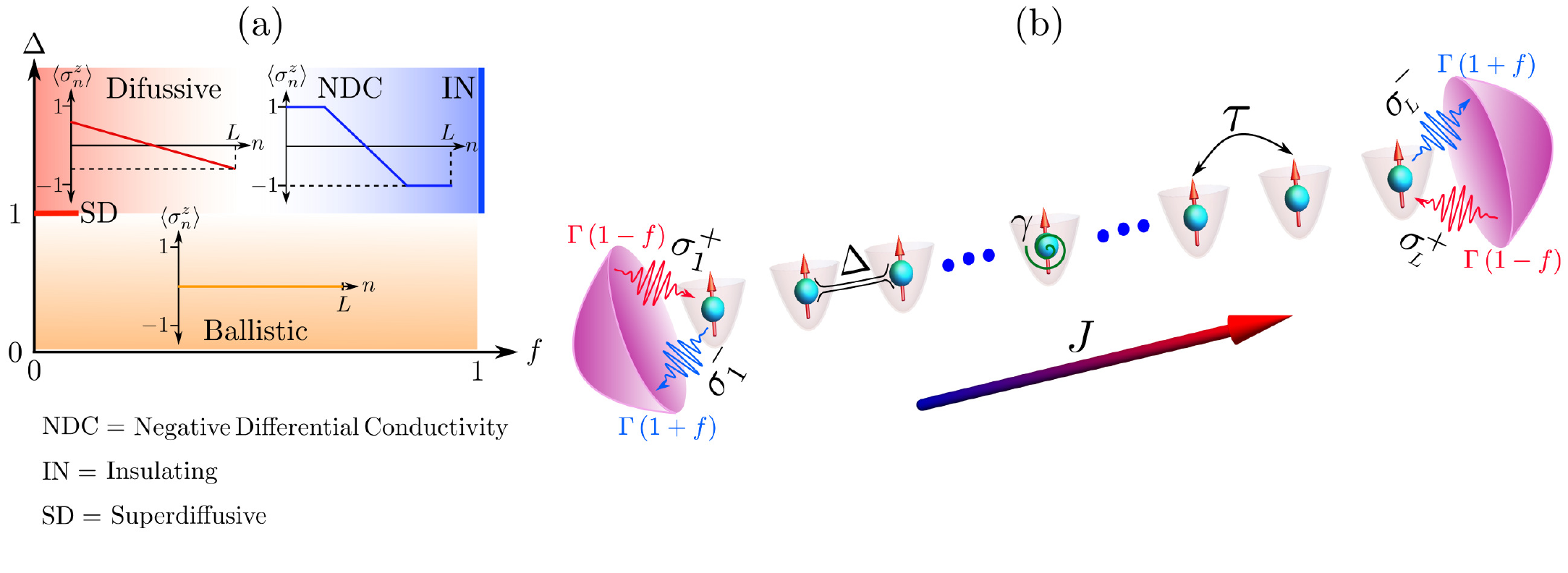}
\caption{Scheme of the system under study. An XXZ spin chain is coupled to unequal reservoirs at its boundaries, which drive it to a NESS where a spin current $J$ flows from left to right. The nonequilibrium $\Delta-f$ phase diagram is also shown, where characteristic magnetization profiles are depicted for each conduction regime. These correspond to a flat profile for ballistic transport, a ramp with homogeneous slope for diffusion, and ferromagnetic domains for the insulating state.}\label{fig_1}
\end{center}
\end{figure}

We consider a spin-$1/2$ chain coupled to two reservoirs of unequal magnetization at its boundaries, as depicted in Fig.~\ref{fig_1}. These induce a net homogeneous spin current in its nonequilibrium steady state (NESS)~\cite{Prosen:2009,Benenti:2009,Mendoza:2013a,Mendoza:2013b,Znidaric:2016,Marko:2017ann}. The chain is characterized by the one-dimensional XXZ Hamiltonian
\beqa\label{Hxxz}
\hat{\mathcal{H}}=\tau\sum_{n=1}^{L-1}\pas{\hat{\sigma}_{n}^{x}\hat{\sigma}_{n+1}^{x}+\hat{\sigma}_{n}^{y}\hat{\sigma}_{n+1}^{y}+\Delta \hat{\sigma}_{n}^{z}\hat{\sigma}_{n+1}^{z}},
\eeqa  
where $\hat{\sigma}_{n}^{\alpha}$ are the Pauli matrices $(\alpha=x,y,z)$, $L$ is the number of sites, $\tau$ is the exchange interaction (we set the energy scale by taking $\tau=1$), and $\Delta$ is the anisotropy along $z$ direction; we take $\Delta\geq0$ without loss of generality. When mapping the model to a spinless-fermion description, $\tau$ corresponds to the hopping and $\tau\Delta$ to a density-density interaction; thus $\Delta>1$ is referred to as the strongly-interacting regime.\\
The boundary reservoirs are modeled as noninteracting spin chains with different average magnetization. In addition, both are weakly coupled to the XXZ chain, and their time correlations decay very rapidly, so their memory effects can be neglected (Markov approximation). Furthermore, the so-called wide-band limit is considered, where the bandwidths of the reservoirs, proportional to their hopping rates, are much larger than the bandwidth of the XXZ chain. Invoking these assumptions~\cite{Benenti:2009,jjthesis}, and following standard microscopic derivations where the degrees of freedom of the reservoirs are traced out~\cite{breuer}, it is shown that the dynamics of the chain under the influence of the environment is governed by a Lindblad master equation,
\begin{equation} \label{lindblad}
\dot{\hat{\rho}}_{t}=\hat{\mathcal{L}}(\hat{\rho})=-i\pas{\hat{\mathcal{H}},\hat{\rho}} + \sum_k\hat{V}_{k}\hat{\rho} \hat{V}_k^{\da}-\frac{1}{2}\pac{\hat{V}_k^{\da}\hat{V}_k,\hat{\rho}}.
\end{equation}
Here $\hat{\rho}$ is the density matrix of the chain, $\hat{\mathcal{L}}$ is the Lindblad superoperator, $\{.,.\}$ is the anticommutator of two operators, and $\hat{V}_k$ correspond to the jump operators establishing the coupling of the chain to the environment. For the boundary driving we consider operators that annihilate $\pap{\hat{V}^{-}}$ and create $\pap{\hat{V}^{+}}$ spin excitations at both the left ($\ell$) and right ({\it r}) edges of the chain, given by $\hat{V}_{{\it r},\ell}^{-}=\sqrt{\Gamma(1\mp f)}\hat{\sigma}_{1,L}^{-}$ and $\hat{V}_{{\it r},\ell}^{+}=\sqrt{\Gamma(1\pm f)}\hat{\sigma}_{1,L}^+$. 
Here $\Gamma$ is the coupling strength (we take $\Gamma=1$), and $\pm f$ are the average magnetizations per spin (in dimensionless units~\cite{Benenti:2009}) of the left/right reservoirs. Thus the parameter $f$ ($0\leq f\leq1$) is the driving, as it establishes the magnetization imbalance between the boundaries. If $f=0$, spin excitations are created and annihilated at the same rate on both boundaries, so there is no net magnetization imbalance and thus no spin transport. If $f>0$, more excitations are created (annihilated) on the left-most (right-most) site of the lattice, inducing a net current resulting from a left-to-right flow and its weaker backflow. We also consider bulk dephasing processes, arising from the coupling of every site of the XXZ chain to local harmonic vibrational degrees of freedom corresponding to the oscillation of the lattice, and quantized in terms of linear phonons~\cite{jjthesis}. Tracing out these degrees of freedom, and considering the wide-band limit, the resulting jump operators are $\hat{V}_{n}^{z}=\sqrt{\gamma}\,\hat{\sigma}_{n}^{z}$ ($n=1,\ldots,L$), where $\gamma$ is the homogeneous dephasing rate.\\
\\
When the spin chain evolves in time as dictated by Eq.~\eqref{lindblad}, it eventually reaches its unique NESS, given by $\hat{\rho}_{\infty} = \lim_{t\to\infty}\hat{\rho}\pap{t}=\lim_{t\to\infty}\exp{\pas{\hat{\mathcal{L}}\pap{t}}}\hat{\rho}\pap{0}$, with $\hat{\rho}\pap{0}$ an initial state. Its transport properties are characterized by its magnetization profile $\langle\hat{\sigma}_{n}^{z}\rangle$ and the homogeneous spin current $\J=\langle \hat{\J}_{n}\rangle$, with $\hat{\J}_{n}=2(\hat{\sigma}_{n}^{x}\hat{\sigma}_{n+1}^{y}-\hat{\sigma}_n^y\hat{\sigma}_{n+1}^{x})$. The phase diagram of the dephasing-free model, obtained from the behavior of both quantities, is depicted in Fig.~\ref{fig_1}. For weak driving $f$ (linear response), $\Delta<1$ shows ballistic conduction while for $\Delta>1$ the system satisfies a normal diffusion equation~\cite{Prosen:2009,Benenti:2009}. This indicates a nonequilibrium quantum phase transition at the isotropic point $\Delta=1$, which is super-diffusive~\cite{Znidaric:2011b}. For large driving $f$ and $\Delta>1$, negative differential conductivity (NDC) emerges, leading to an insulting state at maximal driving $f=1$ in which opposite-polarized ferromagnetic domains suppress transport~\cite{Benenti:2009,Mendoza:2013a,Mendoza:2013b}. This behavior is absent at $\Delta<1$, where transport remains ballistic for any $f$.\\
\\
In the presence of dephasing, spin transport is monotonically degraded with $\gamma$ and becomes diffusive for $\Delta<1$~\cite{Znidaric:2010}, while it is largely enhanced with $\gamma$ for $\Delta>1$, a manifestation of strong correlations~\cite{Mendoza:2013a}. The enhancement vastly increases with driving, as dephasing washes out the NDC and induces an insulator-conducting transition at $f=1$.
\begin{figure}[t]
\begin{center}
\includegraphics[scale=0.8]{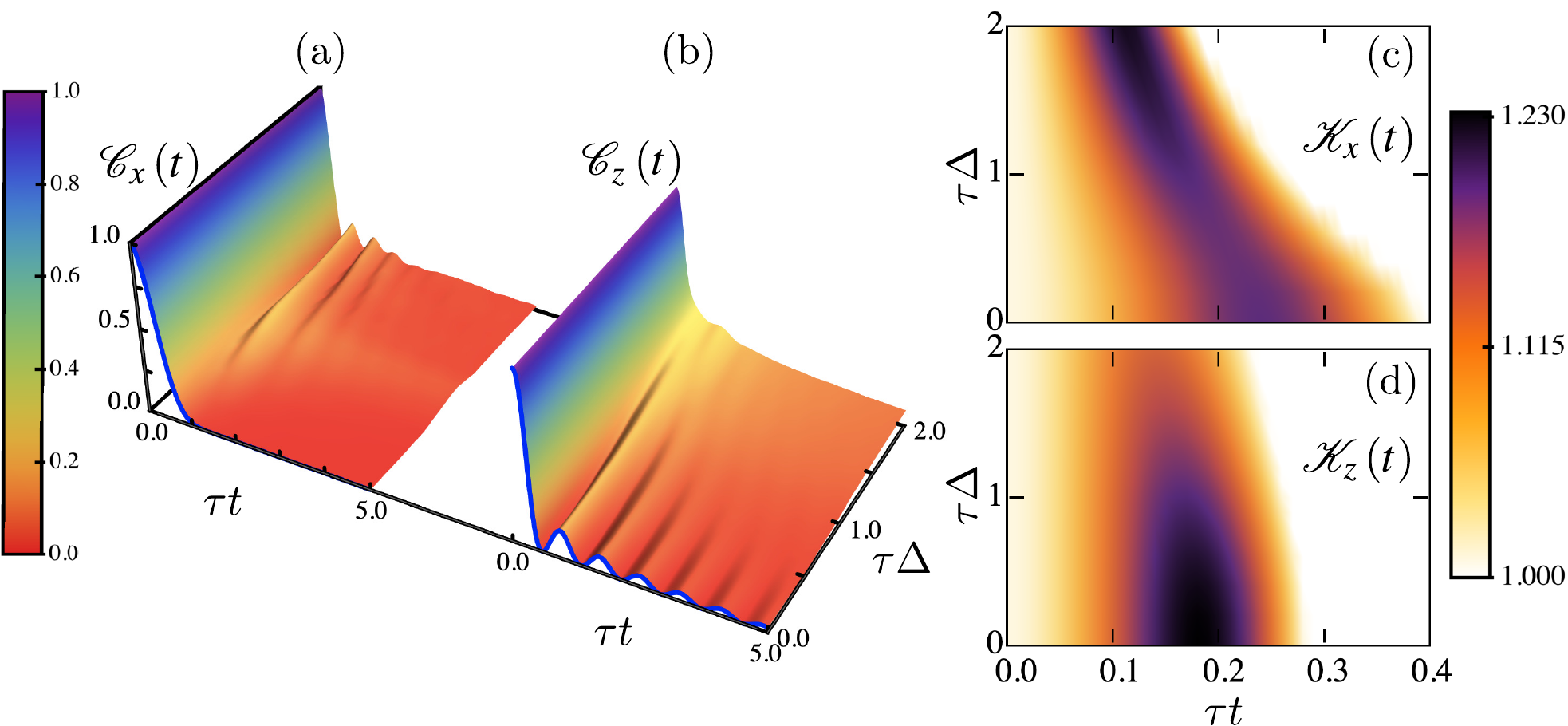}
\caption{Time correlations $\mathcal{C}_{\alpha}(t)$ ((a), (b)) and Leggett-Garg functions $\mathcal{K}^{\alpha}(t)$ ((c), (d)) as a function of time and anisotropy $\Delta$ for weak driving $f=0.1$ and system size $L=60$. (a), (c): $\alpha=x$. (b), (d): $\alpha=z$.  The blue lines in (a) and (b) are the infinite-temperature correlations described in the main text for $\Delta=0$. In (c) and (d), the LGI is violated (satisfied) in the colored (white) regions.}\label{fig_2}
\end{center}
\end{figure}

\section*{Results}
\subsection*{Leggett-Garg Inequalities Calculation}
As shown by Leggett and Garg~\cite{leggett,emary}, macroscopic classical systems satisfy the inequaity $\mathcal{C}\pap{t_1,t_2}+\mathcal{C}\pap{t_2,t_3}-\mathcal{C}\pap{t_1,t_3}\leq 1$ $(t_1<t_2<t_3)$. Here $\mathcal{C}\pap{t_i,t_j}$ denotes the two-time correlations of a dichotomic operator $\hat{Q}$ (with eigenvalues $\pm1$) between times $t_i$ and $t_j$, and is given by $\mathcal{C}\pap{t_i,t_j}=\frac{1}{2}\langle\pac{\hat{Q}\pap{t_i},\hat{Q}\pap{t_j}}\rangle$. For the NESS $\hat{\rho}_{\infty}$, which is unaffected by an evolution under $\hat{\mathcal{L}}$, this correlation with $t_i<t_j$ is given by 
\begin{equation} \label{time_correl}
\mathcal{C}\pap{{\rm t_{i},t_{j}}}={\rm Re}\left(\tr\pas{\hat{Q}\exp{\pas{\hat{\mathcal{L}}\pap{t_j-t_i}}}\hat{Q}\hat{\rho}_{\infty}}\right). 
\end{equation}
Thus, $\mathcal{C}\pap{t_i,t_j}=\mathcal{C}\pap{0,t_j-t_i}$, and taking time intervals $t_2-t_1=t_3-t_2=t$, the LGI reduces to $\mathcal{K}\pap{t}\equiv 2\mathcal{C}\pap{t}-\mathcal{C}\pap{2t}\leq1$, where we define the Leggett-Garg function $\mathcal{K}(t)$, and $\mathcal{C}(t)\equiv \mathcal{C}\pap{0,t}$. Note that we cannot regard the nature of transport as classical from observing that the LGI for some $\hat{Q}$ is satisfied, as it might not capture the quantum correlations that inequalities for other measurements might. However, we do assess genuine quantum behavior when one LGI is violated~\cite{lambert2010prl}.\\ 
\\
We focus on LGIs when performing measurements of local observables $\hat{Q}=\hat{\sigma}_{\lll}^{\alpha}$ ($\alpha=x,z$) for site $\lll=L/2$ on the NESS $\hat{\rho}_{\infty}$. In this form we reduce edge effects as much as possible; however we have verified that the results are qualitatively the same when performing the measurements on different sites (see Fig. S2 of the Supplementary Information (SI)). To evaluate the LGIs violation in the boundary-driven setup for large systems, we obtain their NESS applying the time-dependent density matrix renormalization group~\cite{Zwolak:2004,Cirac:2004}, describing the state by a matrix product structure~\cite{Prosen:2009,Schollwock:2011}. Then we calculate the time evolution of Eq.~\eqref{time_correl} to obtain the single-site two-time correlations, from which the Leggett-Garg function $\mathcal{K}(t)$ follows immediately. Our simulations are based on the open-source Tensor Network Theory (TNT) library~\cite{tnt,tnt_review1}. This constitutes a novel and topical application of matrix product states, fueled by the growing interest in evaluating two-time correlations in dissipative many-body systems~\cite{poletti:2015,everest2017prb,he2017prl,wang2018pre,wolff:2018}.

\subsection*{Controlling LGI violations with interactions: dephasing free case} 
\begin{figure}[t]
\begin{center}
\includegraphics[scale=0.9]{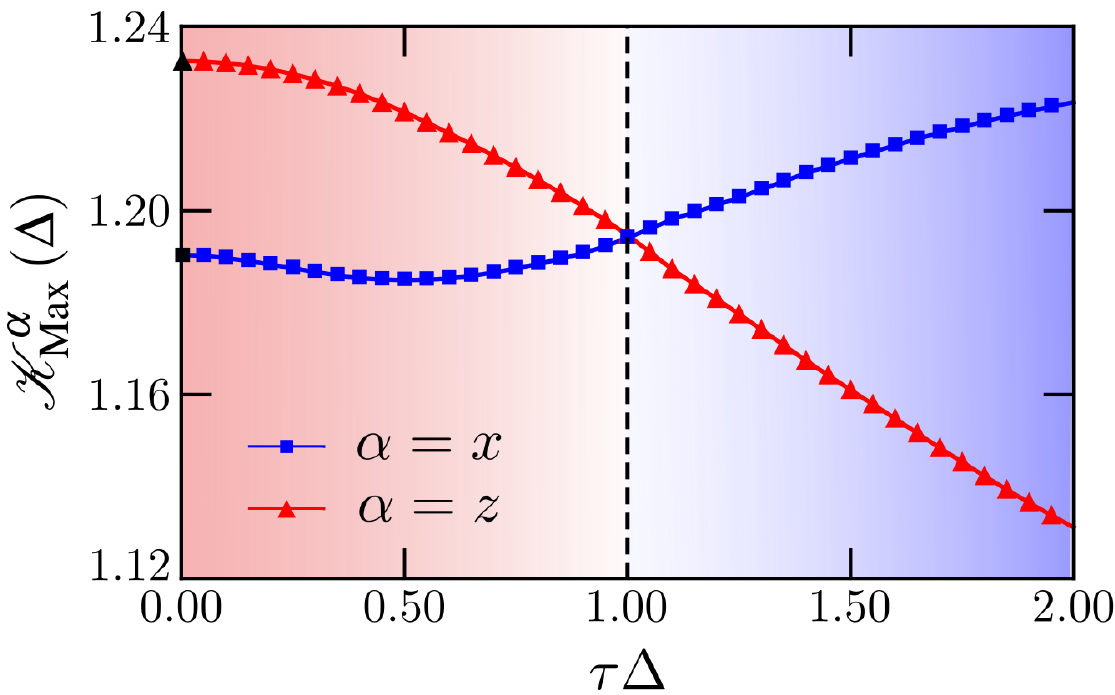}
\caption{Maximum violation of LGI along $x$ and $z$ directions as a function of anisotropy $\Delta$, for weak driving $f=0.1$ and size $L=60$. The black filled symbols at $\Delta=0$ correspond to the infinite-temperature results described in the main text. 
}\label{fig_3}
\end{center}
\end{figure}

We now show that violations of LGIs can be tuned by interactions, and can indicate nonequilibrium quantum phase transitions. For this we consider the linear-response regime, where the system presents a transition from ballistic ($\Delta<1$) to diffusive ($\Delta>1$) spin transport. In Figs.~\ref{fig_2}(a) and~\ref{fig_2}(b) we show $\mathcal{C}_{\alpha}\pap{t}$ for $\alpha=x,z$, which measure the response of the system when adding or removing a spin excitation for $\alpha=x$, or measuring the local magnetization for $\alpha=z$. These correlations already show a qualitative difference, most notable for early times, between the two transport regimes. First, since for $\Delta=0$ and weak driving the NESS is very close to an identity~\cite{Znidaric:2010b,Znidaric:2011a} (with terms of $\mathcal{O}(f)$ accounting for the spin current and magnetization and higher-order corrections for correlations), the results are very well approximated by known analytical results for an infinite-temperature state. These correspond to an exponential decay~\cite{capel:1977} for $\alpha=x$, $\mathcal{C}_x\pap{t}\approx\ee^{-4\tau^2t^2}$, and a squared Bessel function~\cite{katsura:1970} for $\alpha=z$, $\mathcal{C}_z\pap{t}\approx\mathcal{J}_0^2\pap{4\tau t}$, thus being independent of the details of the driving. The qualitative behavior continues when increasing $\Delta$ approximately till the isotropic point. For the strongly interacting regime, the correlations $\mathcal{C}_x\pap{t}$ develop early oscillations of frequency $\sim\Delta^{-1}$, while those of $\mathcal{C}_z\pap{t}$ are strongly suppressed.\\
\\
In Fig.~\ref{fig_2}(c) and~\ref{fig_2}(d) we show the corresponding Leggett-Garg functions; these are obtained exactly for $\Delta=0$, namely $\mathcal{K}^{x}\pap{t}=2\ee^{-4\tau^2t^2}-\ee^{-16\tau^2t^2}$ and $\mathcal{K}^{z}\pap{t}=2\mathcal{J}_0^2\pap{4\tau t}-\mathcal{J}_0^2\pap{8\tau t}$, which agree with our numerical simulations. For all $\Delta$ and early times the inequalities in both directions are violated, indicating genuine quantum behavior which cannot be accounted for by a classical description\footnote{We do not consider longer times, where revivals of the LGI violations occur, as we are mostly interested in the maximal violations $\mathcal{K}_{\text{Max}}^{\alpha}$, which take place at an early time.}. In Fig.~\ref{fig_3} we show the maximal violations $\mathcal{K}_{\text{Max}}^{\alpha}$ as a function of $\Delta$, and observe that while the function of $\alpha=z$ monotonically decreases with interactions, that of $\alpha=x$ has a non-monotonic behavior, but increases for $\Delta>1$. Moreover, at the isotropic point the direction along which the LGI violation becomes maximal changes, from $\alpha=z$ for $\Delta<1$ to $\alpha=x$ for $\Delta>1$. As a result, the direction and magnitude of the overall maximal violation can be controlled by $\Delta$, being enhanced in the strongly-interacting regime. This also shows that the LGI violation indicates the nonequilibrium critical point, a property that it also features in equilibrium~\cite{Gomez:2016prb,Gomez:2018prb}.

\subsection*{Dephasing-enhanced LGI violations} 
\begin{figure}[t]
\begin{center}
\includegraphics[scale=0.7]{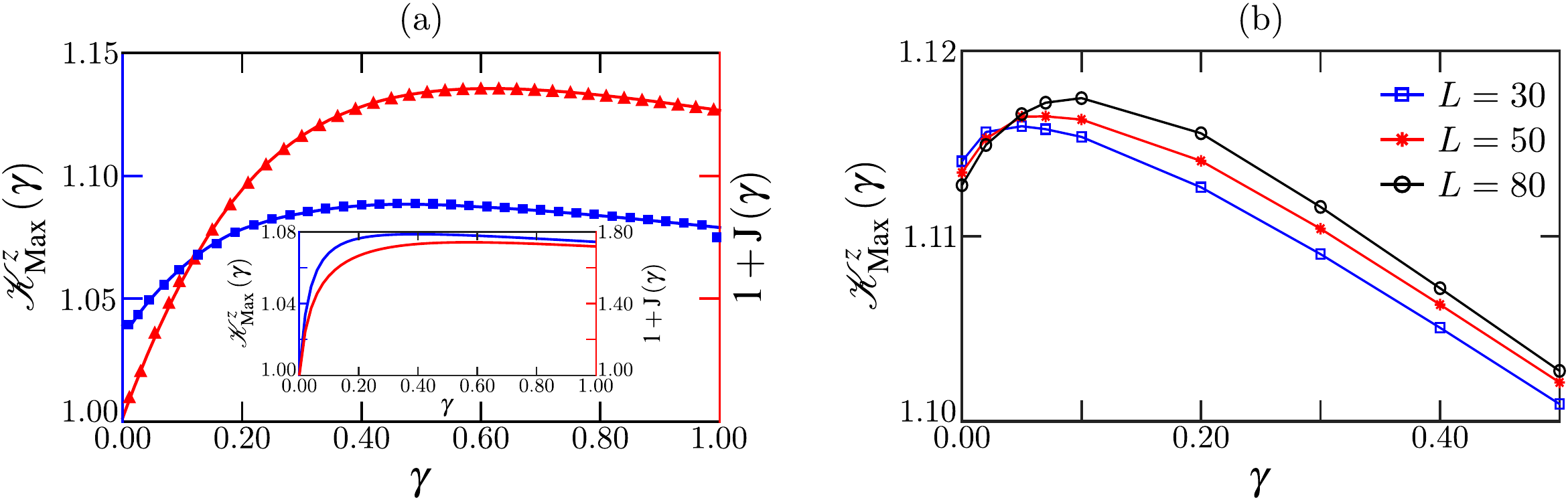}
\caption{(a) Maximum violation of LGI for the XXZ chain along $z$ direction (blue solid line, squares) and shifted spin current (red dashed line, triangles) as a function of dephasing $\gamma$ for $L=16$, anisotropy $\Delta=2$ and driving $f=1$. Inset: Maximal violation (blue solid line) and current (red dashed line) for the minimal model with $f=1$, $K=10$, $\delta=1$ and $\Gamma=1$. (b) Dephasing-enhanced LGI violation at intermediate driving. We depict the maximal LGI violation for $\alpha = z$ as a function of dephasing rate $\gamma$, for $\Delta=2$, $f=0.5$ and different system sizes $L$.}\label{fig_4}
\end{center}
\end{figure}

Now we discuss LGIs of the driven system in the presence of bulk dephasing. As previously mentioned, for strong interactions $\Delta>1$ and large driving $f=1$ dephasing melts the ferromagnetic domains that suppress transport, inducing a transition to a conducting state. Thus the spin current is enhanced with $\gamma$ by several orders of magnitude, as shown in Fig.~\ref{fig_4}(a). This constitutes a many-body scheme of environment-assisted transport~\cite{Mendoza:2013a}. Here we calculate the maximum violation of the LGI for $\alpha=z$ as a function of $\gamma$, also shown in Fig.~\ref{fig_4}(a). Small systems are considered to be able to calculate the NESS in the absence of (or for very weak) dephasing, which takes an exponentially-long time due to the presence of the ferromagnetic domains~\cite{Benenti:2009,Mendoza:2013a}.\\
\\
For $\gamma=0$ LGIs are markedly violated, since operator $\hat{Q}$ is applied between the ferromagnetic domains and induces appreciable dynamics. But remarkably, as the system becomes conducting due to weak dephasing, it also presents enhanced quantumness measure manifested in the increase of the violations. Even for stronger dephasing $\gamma\approx1$, where $\mathcal{K}^z_{\text{Max}}$ decreases with $\gamma$, the violation is still larger than that of $\gamma=0$. Thus there is a wide range of environmental couplings where in addition to a large transport enhancement, quantum features are strengthened. This is in stark contrast to the weakly-interacting case, depicted in Fig. S1 of the SI, where both the spin transport and the violation of LGIs are monotonically degraded with dephasing. This indicates that the origin of the observed effect is truly a consequence of strong interactions between particles, a scenario where LGIs remain largely unexplored.

We also observe enhanced violation of LGIs with dephasing even for much weaker driving $f$, where for $\gamma=0$ the system is conducting and the differential conductivity is positive (as NDC emerges at larger driving)~\cite{Benenti:2009,Mendoza:2013a}. In Fig.~\ref{fig_4}(b) we present the maximal violation as a function of $\gamma$ for $\Delta=2$ and an intermediate driving $f=0.5$, in which the system is conducting and far away from the insulating limit. We first note that in the absence of dephasing, the violation for $f=0.5$ is larger than that at $f=1$ (see Fig.~\ref{fig_4}(a)), which shows that just making the system conducting by inducing a magnetization backflow increases the quantumness measure in the lattice. Furthermore, when dephasing is included, it increases the violation compared to the $\gamma=0$ case; this indicates that the effect is not just a peculiarity of the maximally-driven limit, but emerges in a broader transport regime. The key ingredient is again the existence of strong interactions $\Delta$. This strongly contrasts with the natural expectation that in conducting regimes, environmental coupling degrades LGI violations~\cite{lambert2010prl,robens2015prx}.\\
\\
It is important to stress two additional properties evidenced in Fig.~\ref{fig_4}(b). First, the maximal violation barely changes with the system size, indicating that this is not a small-lattice effect. We have performed simulations around the optimal dephasing rate for systems of up to $L=150$, and observed that $\mathcal{K}_{\text{Max}}^z$ slightly increases with $L$, from $\mathcal{K}_{\text{Max}}^z=1.116$ for $L=30$ to $\mathcal{K}_{\text{Max}}^z=1.120$ (not shown). Thus for large systems the positive impact of dephasing is still present and of almost the same amplitude of that at much smaller lattices. Second, the optimal dephasing slightly shifts to higher values with $L$, but for the largest chains considered, it remains $\approx0.10$. This leads to a wide range of dephasing rates for which the LGI violation is larger than that of the $\gamma=0$ limit.\\
\\
Finally, we note that the LGI violation enhancement by dephasing is also observed when non-local observables $\hat{Q}$ are considered. In particular, we calculated the two-time correlations for strings of operators $\hat{Q}=\hat{\sigma}_{\lll}^{z}\hat{\sigma}_{\lll+1}^{z}\hat{\sigma}_{\lll+2}^{z}\cdots\hat{\sigma}_{n}^{z}$ of different length, located around the center of the XXZ lattice. The results, shown of Fig. S3 of the SI, are very similar to those of Fig.~\ref{fig_4}, both at and away from strong driving, where the maximal LGI violation takes place at a finite dephasing rate. This makes our conclusions more robust, and suggests that this phenomenon could be observed in experiments where measurements are performed on a non-local basis.

\subsection*{Minimal model} 
\begin{figure}[h!]
\begin{center}
\includegraphics[scale=0.8]{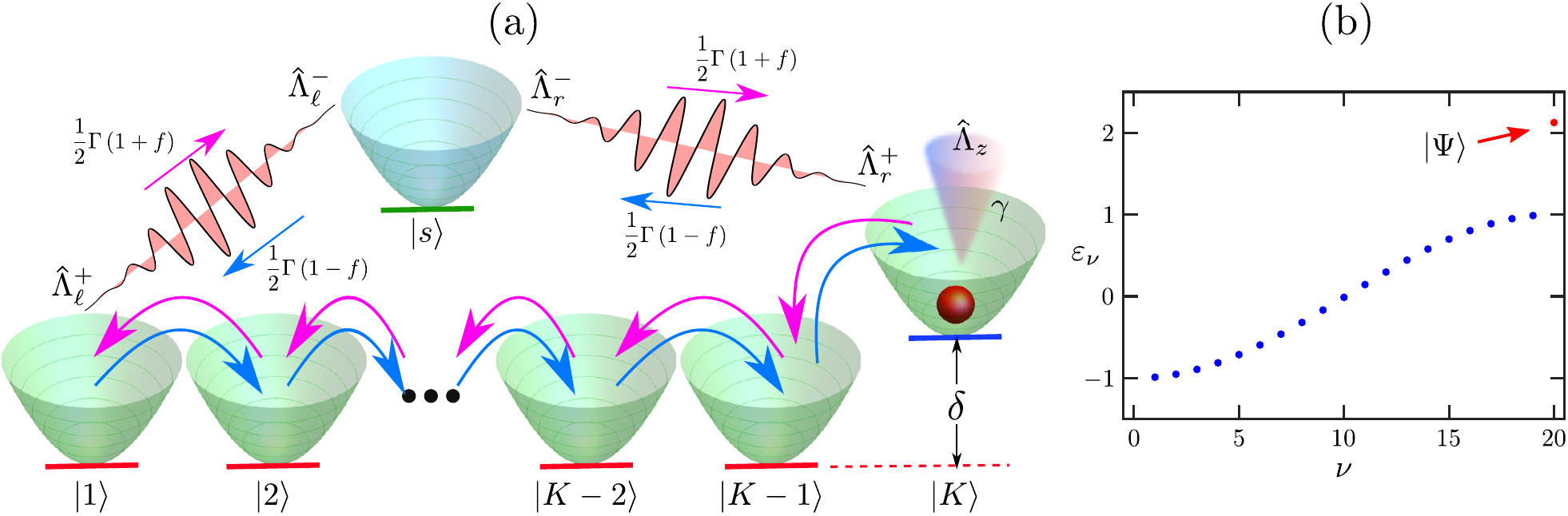}
\caption{Minimal model. (a) Schematic illustration. A single-particle system is incoherently coupled to an auxiliary state $|s\rangle$, which induces a current with driving $f$. (b) Eigenstructure. The depicted energies $\varepsilon_{\nu}$ correspond to $K=20$ and $\delta=2$. The gap between the most energetic eigenstate $|\Psi\rangle$ (red dot), preferentially populated at maximal driving and zero dephasing (see Eq. (S1) of the SI), and the conducting band, is of $\mathcal{O}(\delta)$.}\label{fig_3_sm}
\end{center}
\end{figure}
To obtain more insight into the incoherent enhancement of LGI violations, we analyze a simple model which incorporates the essential ingredients of the strongly-interacting lattice~\cite{Mendoza:2013a}. This model, depicted in Fig.~\ref{fig_3_sm}, mimics its gapped eigenstructure, consisting of flat (insulating) and wide (conducting) bands, and the population transfer between them due to coherent processes, incoherent driving and dephasing. It consists of $K$ levels $|1\rangle,\ldots,|K\rangle$, with Hamiltonian
\begin{equation}
\hat{\mathcal{H}}_{\text{Min}}=\frac{1}{2}\sum_{k=1}^{K-1}\left(|k\rangle\langle k+1|+\text{H.c.}\right)+\delta|K\rangle\langle K|,
\end{equation} 
where state $|K\rangle$ is elevated in energy with respect to the others by an amount $\delta>0$; thus the bands are separated by a gap of $\mathcal{O}(\delta)$, similarly to the strongly-interacting XXZ model (gaps of $\mathcal{O}(\Delta)$). The driving is modeled by the incoherent coupling of states $|1\rangle$ and $|K\rangle$ to an auxiliary site $|s\rangle$, with jump operators $\hat{V}_{\ell,r}^{-}=\sqrt{\Gamma(1\mp f)}\hat{\Lambda}_{\ell,r}^{-}$ and $\hat{V}_{\ell,r}^{+}=\sqrt{\Gamma(1\pm f)}\hat{\Lambda}_{\ell,r}^+$, where $\hat{\Lambda}_{\ell}^{-}=|s\rangle\langle1|$, $\hat{\Lambda}_{r}^{+}=|K\rangle\langle s|$, $\hat{\Lambda}_{\ell}^{+}=(\hat{\Lambda}_{\ell}^{-})^{\dagger}$ and $\hat{\Lambda}_{r}^{-}=(\hat{\Lambda}_{r}^{+})^{\dagger}$. In addition, the dephasing at rate $\gamma$ corresponds to the jump operator $\hat{\Lambda}_{z}=\sqrt{\gamma}(\hat{I}-2|K\rangle\langle K|)$, with identity $\hat{I}$, which is equivalent to $\hat{\sigma}^z$ in the XXZ model and acts only on the state of highest energy. This is similar to the effect of dephasing on the original many-body system, which induces transitions from high-energy flat bands of the eigenstructure to low-energy wide bands of higher conduction~\cite{Mendoza:2013a}. Finally, the associated local current operator is $\hat{\J}_{\text{Min},k}=-i\left(|k\rangle\langle k+1|-\text{H.c.}\right)$.\\
\\
This model leads to the emergence of NCD, an insulating state at $f=1$ and dephasing-enhanced transport. These effects are shown in Fig. \ref{fig_4_sm}(a) for $K=10$, $\Gamma=1$ and $\delta=1$. Note that $\delta=1$ does not correspond to the isotropic point of the XXZ model. The model does not capture the transport phase transition, as it is defined to mimic the strongly-interacting regime of the spin chain. Here $\delta$ sets the size of the gap between bands~\cite{Mendoza:2013a}. We clearly see that when $\gamma=0$ the system is insulating at maximal driving $f=1$, where the equivalent of the ferromagnetic domains is a preferred population of state $|K\rangle$ (see Eq. (S1) of the SI). The NDC is washed out by dephasing, which results in a large transport enhancement with $\gamma$. For large dephasing the transport is degraded, as a result of the Zeno effect which tends to freeze the dynamics. Thus the main transport properties of the strongly-interacting XXZ system are captured by the minimal model. 
\begin{figure}[t]
\begin{center}
\includegraphics[scale=0.65]{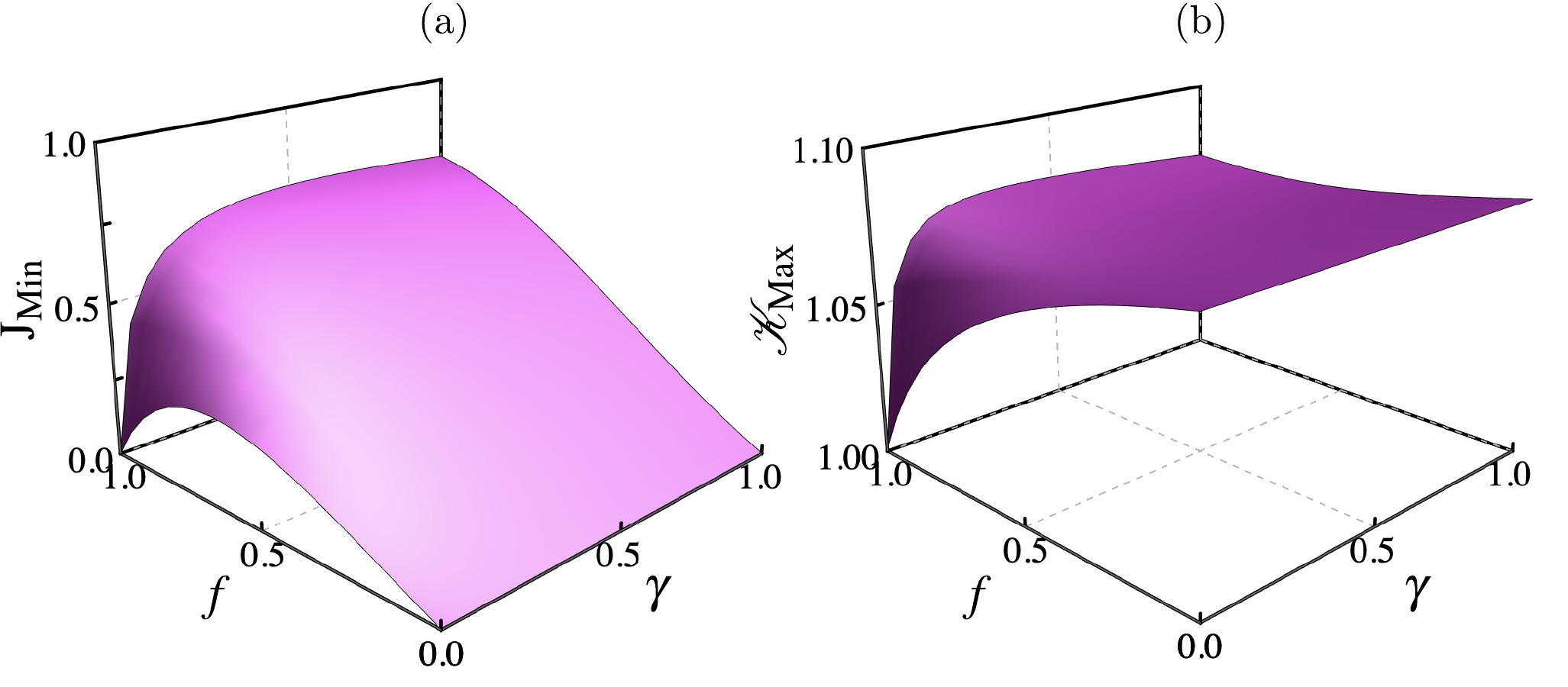}
\caption{Physical properties of the minimal model. (a) Current $\J_{\text{Min}}=\langle\hat{\J}_{\text{Min},k}\rangle$ as a function of driving $f$ and dephasing $\gamma$. (b) Corresponding maximal value of the Leggett-Garg function, i.e. maximal violation of LGI.}\label{fig_4_sm}
\end{center}
\end{figure}
To study LGIs, we calculate the time correlations for the operator $\hat{Q}=\hat{\mathcal{I}}-2|P\rangle\langle P|$, where site $P$ is taken on the central site of the system. Exact numerical results of the Leggett-Garg function $\mathcal{K}(t)$ are depicted in Fig.~\ref{fig_4_sm}(b), showing that, similarly to the original model, two mechanisms enhance the violations of LGIs with respect to the dephasing-free $f=1$ limit (where there is no visible violation), namely introducing a backflow ($f<1$) or dephasing ($\gamma>0$). Very large dephasing degrades this scenario, as expected. Thus in spite of its simplicity, the model features the essential ingredients underlying such phenomena.\\
\\
Analytical results up to $\mathcal{O}(\delta^{-2})$ for the strongly-driven limit, where the effects of decoherence are the largest, help determine how LGI violations are increased; see details in the SI. For the dephasing-free system with $f=1$, $\mathcal{C}^{(0)}(t) \approx \mathcal{K}^{(0)}(t) \approx1$, so the LGI is not significantly violated, as expected. Defining $\beta=2(1-f)(1+\Gamma^2)$ when slightly moving from maximal driving with $\gamma=0$, or $\beta=16\gamma(1+\Gamma^2)/\Gamma$ when introducing weak dephasing at $f=1$, or their sum if both are present, we get the general behavior
\begin{equation} \label{lgi_f}
\mathcal{K}(t)=1+\beta\left(\frac{t}{2\delta}\right)^2.
\end{equation}
This shows that both mechanisms induce an appreciable violation of the LGI which grows quadratically at early times, and which increases linearly as $(1-f)$ and/or $\gamma$. It also manifests that the source of these results, also present in the strongly-interacting XXZ model, is the gapped eigenstructure of the model.

\section*{Discussion} 
We have discussed schemes for increasing violations of Leggett-Garg inequalities in open many-body systems, illustrated in archetypical nonequilibrium boundary-driven configurations. Using matrix product state simulations, we have determined that in the strongly-interacting regime, interactions and (unexpectedly) bulk dephasing can enhance the violations. The main mechanism behind the latter effect is illustrated by the same observation in a minimal model, which mimics the gapped eigenstructure of the original system. Given the simplicity of the studied models, our results pave the way for novel experimental protocols of detecting macroscopic quantum coherence in a large variety of scenarios, including single-particle and many-body inhomogeneous (e.g. disordered~\cite{Znidaric:2016,Marko:2017ann}) current-carrying systems.

\bibliography{LGI_NESS_Bib}{}
\bibliographystyle{apsrev4-1}
\section*{Acknowledgements}
J.J.M.-A. thanks Stephen Clark and Katarzyna Roszak for interesting discussions. We acknowledge financial support from Facultad de Ciencias-UniAndes-2019 project \emph{Quantum thermalization and optimal control in many-body systems} (Grant No. INV-2018-50-1384). We acknowledge financial support from Departamento Administrativo de
Ciencia, Tecnolog\'ia e Innovaci\'on Colciencias, through the project \emph{Producci\'on y Caracterizaci\'on de Nuevos Materiales Cu\'anticos de Baja Dimensionalidad: Criticalidad Cu\'antica y Transiciones de Fase Electr\'onicas} (Grant No. 120480863414). 
\section*{Author contributions}	
J. J. Mendoza-Arenas developed the theoretical framework and implemented the time evolution codes. F. J. Gómez-Ruiz developed numerical simulations and prepared the figures. All authors contributed to the analysis of the results and the writing of the manuscript. L. Quiroga and F. J. Rodríguez supervised the project.
\newpage
\pagebreak
\clearpage
\setcounter{equation}{0}
\setcounter{figure}{0}
\setcounter{table}{0}
\setcounter{section}{0}
\setcounter{page}{1}
\makeatletter
\renewcommand{\thesection}{S\arabic{section}}
\renewcommand{\theequation}{S\arabic{equation}}
\renewcommand{\thefigure}{S\arabic{figure}}
\renewcommand{\bibnumfmt}[1]{[S#1]}
\renewcommand{\citenumfont}[1]{S#1}
\newcounter{fnnumber}
\renewcommand{\thefootnote}{\fnsymbol{footnote}}
\begin{center}
\textbf{\large ---Supplemental Material---\\
Enhancing violations of Leggett-Garg inequalities in nonequilibrium correlated many-body systems by interactions and decoherence}\\
\vspace{0.5cm}
J. J. Mendoza-Arenas,$^{1}$ F. J. G\'omez-Ruiz,$^{2,1,}$\footnote{Electronic address: \href{mailto:fj.gomez34@dipc.org}{fj.gomez34@dipc.org}}\: F. J. Rodr\'iguez,$^{1}$ and L. Quiroga$^{1}$\\
\vspace{0.2cm}
$^1${\it Departamento de Física, Universidad de los Andes, A.A. 4976, Bogotá D. C., Colombia.}\\
$^{2}${\it Donostia International Physics Center,  E-20018 San Sebasti\'an, Spain}
\end{center}
\begin{center}
\begin{tabular}{p{14cm}}
\vspace{0.1cm}
\quad In this Supplementary Information (SI), we provide details of the analytical strategies employed in the main text to obtain exact numerical results for two-time correlations and Leggett-Garg inequalities for a nonequilibrium setup of quantum transport.
\end{tabular}
\end{center}
\section{Weakly-interacting system}

\begin{figure}[h!]
\begin{center}
\includegraphics[scale=0.8]{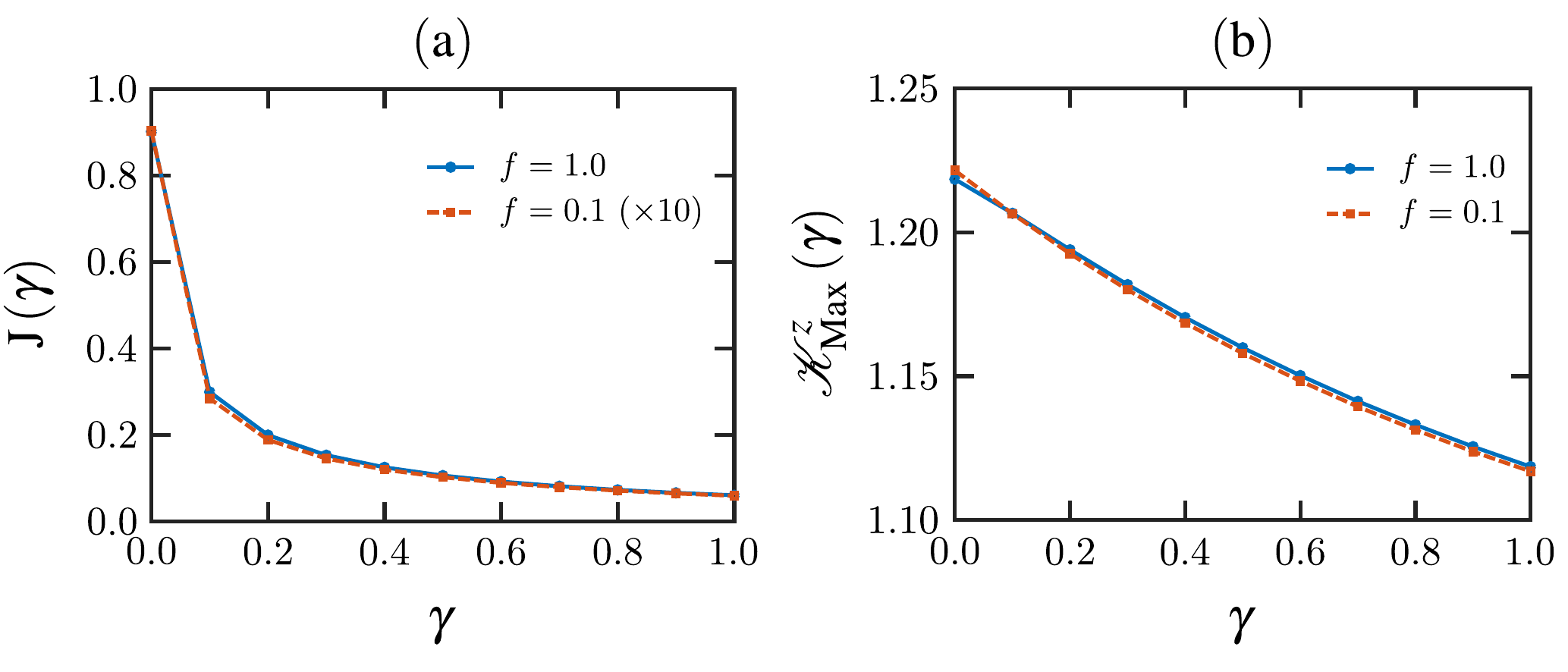}
\caption{{\bf Weakly-interacting XXZ model.} (a) Spin current as a function of dephasing, for $L=60$ sites, $\Delta=0.5$, $\alpha=z$ and both weak ($f=0.1$) and strong ($f=1.0$) driving. (b) Corresponding maximal value of Leggett-Garg functions.}\label{fig_1_sm}
\end{center}
\end{figure}

In this section we briefly illustrate the impact of bulk dephasing on LGIs in the weakly-interacting regime ($\Delta<1$) of the nonequilibrium model. In Fig.~\ref{fig_1_sm} we show how, similarly to spin transport, the value of the Leggett-Garg function for $\alpha=z$ monotonically decreases with the dephasing rate $\gamma$. Thus the quantumness of the transport supported by the system is degraded by environmental coupling, as intuitively expected~\cite{lambert2010prl}. It is worth noting that the inequalities are violated even for the largest values of dephasing considered, so the transport is of quantum nature and its properties cannot be accounted for classically.

\section{LGI violations for measurements on different sites}

\begin{figure}[h!]
\begin{center}
\includegraphics[scale=0.7]{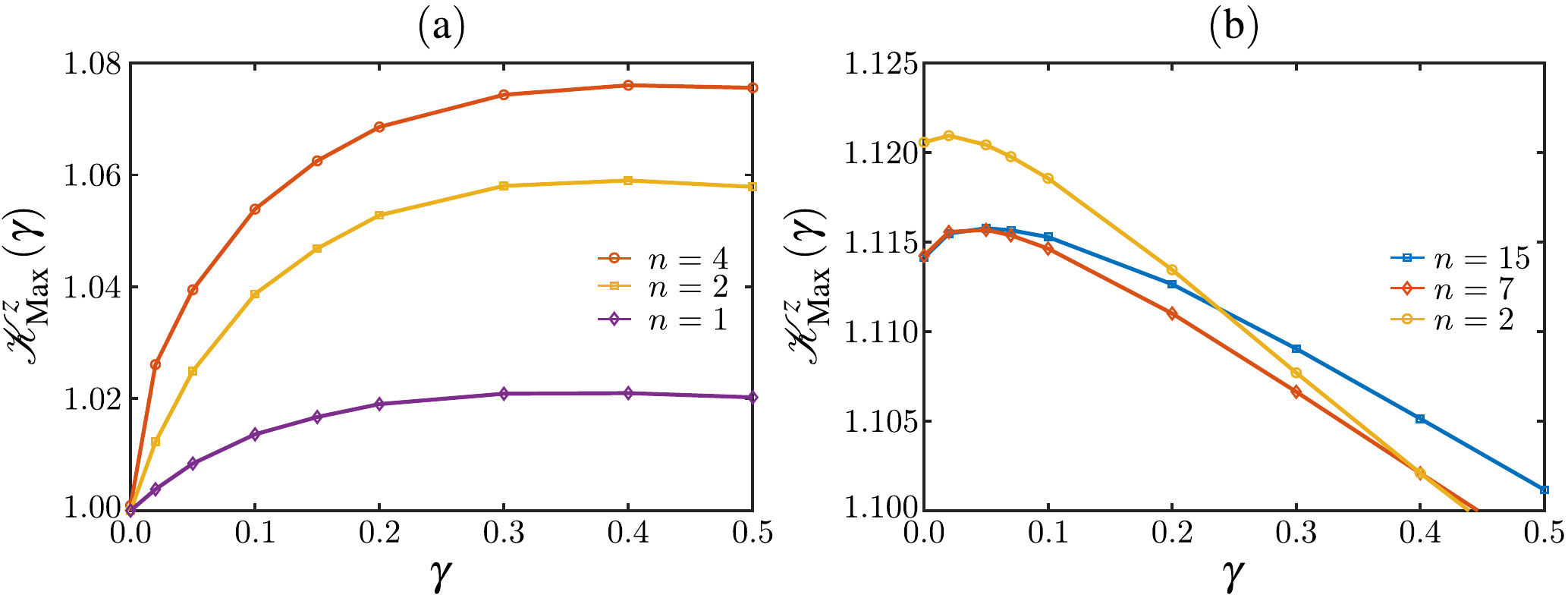}
\caption{{\bf Dephasing-enhanced LGI violation for different sites $n$ where correlations are measured.} (a) At maximal driving $f=1$ and $L=16$. (b) At intermediate driving $f=0.5$ and $L=30$. We depict the maximal LGI violation for $\alpha = z$ as a function of dephasing rate $\gamma$, for $\Delta=2$.}\label{fig_2_sm}
\end{center}
\end{figure}
In this section we discuss how the main result of our work, namely the dephasing-enhanced LGI violation for strong interactions, is modified when we consider observables different to $\hat{\sigma}^{z}_{L/2}$. First we evaluate the two-time correlations for local operators applied to different sites. The results are shown in Fig.~\ref{fig_2_sm}. For maximal driving $f=1$, the Leggett-Garg function has a very similar behavior to that evaluated at the central site (Fig. 4(a) of the main text), but with a lower violation since excitations created closer to the boundaries propagate more slowly. This is more clearly seen for weak dephasing, due to the presence of the ferromagnetic domains. In fact, for zero dephasing, no violation is seen for measurement sites away from the centre. For intermediate driving, deep inside the chain, the Leggett-Garg function is essentially site-independent for low dephasing rates $\gamma$; differences emerge for larger $\gamma$, for which the violation of LGIs starts to be degraded. For sites closer to the boundary the violation enhancement compared to $\gamma=0$ is weaker, but still notable. Thus the main results of violation enhancement by dephasing are not exclusive to the central spin.\\
\\
Now we consider the effect of non-local measurements. For this we take strings of operators $\hat{Q}=\hat{\sigma}_l^{z}\hat{\sigma}_{l+1}^{z}\hat{\sigma}_{l+2}^{z}\cdots\hat{\sigma}_n^{z}$ of different length, located around the central site of the lattice. The results, shown in Fig.~\ref{fig_3_sm}, are qualitatively very similar to those obtained for the central site (and different sites, as seen in Fig.~\ref{fig_2_sm}). For maximal driving we again see an enhancement of the violation even for large values of $\gamma$. However, as the string of $\hat{\sigma}_l^{z}$ operators includes sites closer to the boundaries, the maximal violation decreases; furthermore, for zero dephasing, no appreciable violation is seen for long-enough strings. For intermediate driving, the violation of the LGIs for several sites is also similar to that of a single site, with a very small but observable enhancement by dephasing. These results indicate that our observations remain qualitatively unaffected if non-local operators are considered instead of single-site operators.

\begin{figure}[h!]
\begin{center}
\includegraphics[scale=0.7]{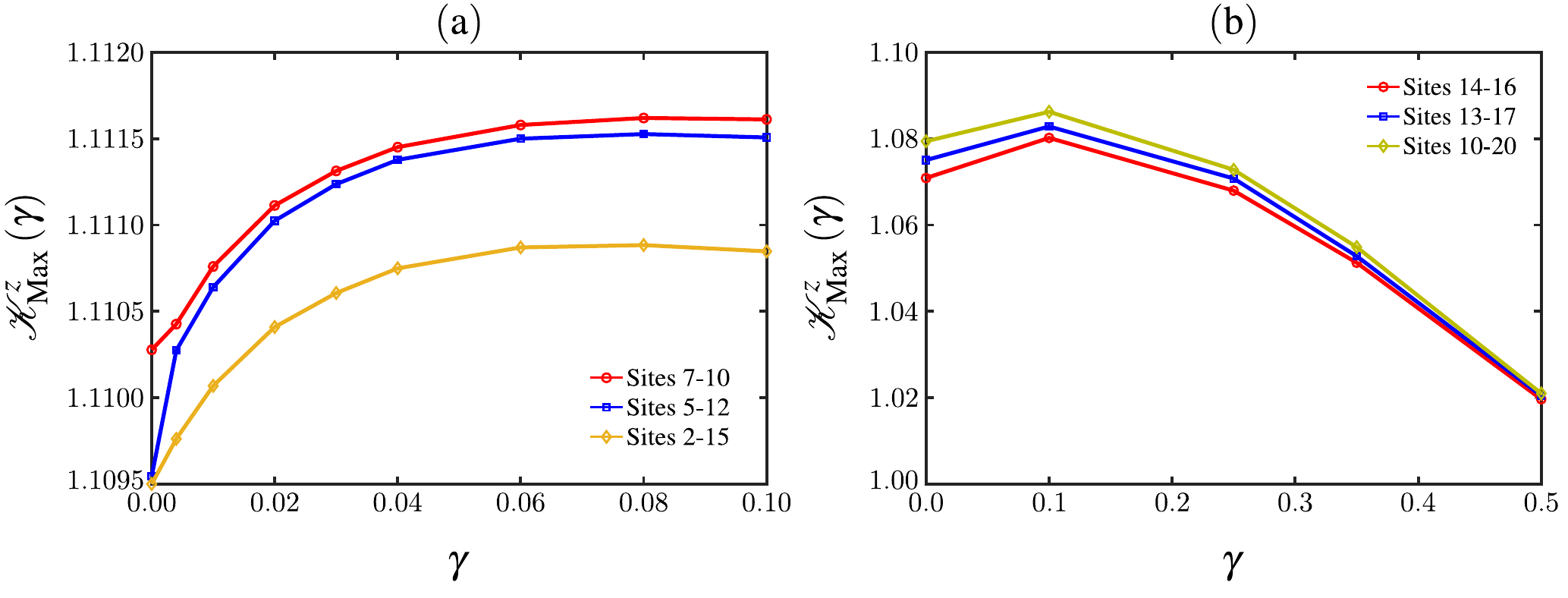}
\caption{{\bf Dephasing-enhanced LGI violation for operators acting on several sites.} (a) At maximal driving $f=1$ and $L=16$. (b) At intermediate driving $f=0.5$ and $L=30$. We depict the maximal LGI violation for $\alpha = z$ as a function of dephasing rate $\gamma$, for $\Delta=2$.}\label{fig_3_sm}
\end{center}
\end{figure}

\section{Analytic results on the minimal model} \label{appendix_toy}
In the following we describe an analytic approach to obtain the Leggett-Garg functions of the simple minimal model, focused on large driving and weak dephasing. We first note that in the maximally-driven limit $f=1$ and in the absence of dephasing, this model features an insulating NESS akin to that of the original model of the main text, of the form~\cite{Mendoza:2013a}
\begin{align} \label{ness_toy_f1}
\hat{\rho}^{(0)}=\frac{1}{\mathcal{N}}|\Psi\rangle\langle\Psi|,\quad|\Psi\rangle=\sum_{n=0}^{K-1}(2\delta)^{-n}|K-n\rangle,
\end{align} 
with normalization constant
\begin{equation}
\mathcal{N}=\frac{4\delta^2-(4\delta^2)^{1-K}}{4\delta^2-1}.
\end{equation}
This result shows that the population of a site exponentially decreases as it gets far from the rightmost site $|K\rangle$.

\subsection{Perturbative $1-f$ solution to the minimal model} \label{appendix_toy_pert}
Now we move from the maximally-driven case, and consider the situation where a weak back-flow is introduced, while keeping $\gamma=0$. We obtain the correction to the NESS to first order in $\mu=1-f\ll1$, given by
\begin{equation} \label{pertur_mu}
\hat{\rho}_{\infty}=\hat{\rho}^{(0)}+\mu\hat{\rho}^{(1)}.
\end{equation} 
For this we rewrite the Lindblad superoperator of Eq. (2) of the main text in the form
\begin{equation} \label{lindblad_toy}
\hat{\mathcal{L}}(\hat{\rho})=\hat{\mathcal{L}}^{(0)}(\hat{\rho})+\mu\hat{\mathcal{L}}^{(1)}(\hat{\rho}),
\end{equation}
where $\hat{\mathcal{L}}^{(0)}(\hat{\rho})$ corresponds to the $\mu=0$ Lindblad superoperator, namely
\begin{equation}
\hat{\mathcal{L}}^{(0)}\pap{\hat{\rho}}=-\ii\pas{\hat{\mathcal{H}},\hat{\rho}}+\Gamma\left(\mathcal{D}_{\hat{\Lambda}_{\ell}^{-}}+\mathcal{D}_{\hat{\Lambda}_{r}^{+}}\right),
\end{equation}
and $\hat{\mathcal{L}}^{(1)}(\hat{\rho})$ to the remaining terms,
\begin{equation}
\hat{\mathcal{L}}^{(1)}\pap{\hat{\rho}}=\frac{\Gamma}{2}\left(\mathcal{D}_{\hat{\Lambda}_{\ell}^{+}}+\mathcal{D}_{\hat{\Lambda}_{r}^{-}}-\mathcal{D}_{\hat{\Lambda}_{\ell}^{-}}-\mathcal{D}_{\hat{\Lambda}_{r}^{+}}\right),
\end{equation}
where each dissipator $\mathcal{D}$ is given by
\begin{equation}
\mathcal{D}_{\hat{X}}(\hat{\rho})=\hat{X}\hat{\rho}\hat{X}^{\dagger}-\frac{1}{2}\{\hat{X}^{\dagger}\hat{X},\hat{\rho}\}.
\end{equation}
To solve Eq.~\eqref{lindblad_toy} for the NESS $\hat{\rho}_{\infty}$, we use the anstaz
\begin{equation}
\hat{\rho}_{\infty}=\sum_{p=0}^{\infty}\mu^{p}\hat{\rho}^{(p)}.
\end{equation}
Gathering terms of equal powers of $\mu$, we obtain the equations that determine the different order corrections. For $p=0$, this corresponds to $\hat{\mathcal{L}}^{(0)}\pap{\hat{\rho}^{(0)}}=0$, which is the $\mu=0$ equation with the solution of Eq.~\eqref{ness_toy_f1}. For $p>0$ we have
\begin{equation} \label{pertur_p_order}
\hat{\mathcal{L}}^{(0)}(\hat{\rho}^{(p)})+\hat{\mathcal{L}}^{(1)}(\hat{\rho}^{(p-1)})=0,
\end{equation}
which indicates that to obtain the correction of $\mathcal{O}(p)$, that of $\mathcal{O}(p-1)$ is required. Importantly, since $\hat{\rho}^{(0)}$ is a valid density matrix, its trace is one, so the $p>0$ corrections are traceless. In the following calculation we restrict to $p=1$, valid for a weak deviation from the maximally-driven system. Equation~\eqref{pertur_p_order} becomes
\begin{equation} \label{pertur_1_order}
\hat{\mathcal{L}}^{(0)}(\hat{\rho}^{(1)})+\hat{\mathcal{L}}^{(1)}(\hat{\rho}^{(0)})=0,
\end{equation}
from which we can obtain $\hat{\rho}^{(1)}$ from $\hat{\rho}^{(0)}$. First we calculate $\hat{\mathcal{L}}^{(1)}(\hat{\rho}^{(0)})$, obtaining
\begin{equation} \label{L1rho0}
\begin{split}
\hat{\mathcal{L}}^{(1)}(\hat{\rho}^{(0)})=\frac{\Gamma}{2\mathcal{N}}\biggl\{&|s\rangle\langle s|\left(1+(2\delta)^{-2(K-1)}\right)-\frac{1}{2}\sum_{n=0}^{K-1}(2\delta)^{-n}(|K\rangle\langle K-n|+\text{H.c.})\\
&-\frac{1}{2}\sum_{n=0}^{K-1}(2\delta)^{-(K-1+n)}(|1\rangle\langle K-n|+\text{H.c.})\biggr\}.
\end{split}
\end{equation} 
To calculate $\hat{\mathcal{L}}^{(0)}(\hat{\rho}^{(1)})$ we consider the following general form for $\hat{\rho}^{(1)}$,
\begin{equation}
\hat{\rho}^{(1)}=a_s| s\rangle+\sum_{m,n=0}^{K-1}a_{m,n}|K-m\rangle\langle K-n|.
\end{equation}
Note that site $|s\rangle$ has no coherences with other sites, as it is only incoherently coupled to other sites ($|1\rangle$ and $|K\rangle$). We obtain
\begin{align} \label{L0rho1}
\begin{split}
\hat{\mathcal{L}}^{(0)}(\hat{\rho}^{(1)})&=\Gamma a_s |K\rangle\langle K|+\Gamma(a_{K-1,K-1}-a_s)|s\rangle\langle s|-\frac{\Gamma}{2}\sum_{m=0}^{K-1}(a_{K-1,m}|1\rangle\langle K-m|+\text{H.c.})\\
&+\frac{1}{2}\sum_{m=0}^{K-1}\sum_{n=1}^{K-1}(ia_{m,n}|K-m\rangle\langle K-n+1|+ia_{m,n-1}|K-m\rangle\langle K-n|)+\text{H.c.})\\
&+\delta\sum_{m=1}^{K-1}(ia_{n,0}|K-n\rangle\langle K|+\text{H.c.}).
\end{split}
\end{align}
After gathering the coefficients $a_{m,n}$ of all the different basis elements $|K-m\rangle\langle K-n|$ in Eqs.~\eqref{L1rho0} and~\eqref{L0rho1}, and considering that $a_{m,n}^*=a_{n,m}$ so the density matrix is Hermitian, we obtain a linear system of equations whose solution gives the correction $\hat{\rho}^{(1)}$. Rewriting each correction coefficient as $a_{m,n}=r_{m,n}+ij_{m,n}$ and considering the Hermiticity of the density matrix ($r_{m,n}=r_{n,m}$ and $j_{m,n}=-j_{n,m}$), this is transformed into a linear system of $1+K^2$ equations with $(K^2+K+2)/2$ unknowns $r_{m,n}$ (including $r_s\equiv a_s$) and $(K^2-K)/2$ unknowns $j_{m,n}$ (the diagonal elements must be real). In addition, since $\hat{\rho}^{(1)}$ is traceless, the diagonal coefficients must also satisfy
\begin{equation}
r_s+\sum_{n=0}^{K-1}r_{n,n}=0.
\end{equation}
We solve this system of equations up to $\mathcal{O}(\delta^{-2})$, for which $\hat{\rho}^{(1)}$ has the (almost tridiagonal) form
\begin{equation}
\hat{\rho}^{(1)}=\begin{pmatrix}
r_{K-1} & -ij_1 & & & \cdots &&&&& \\
ij_1 & r_b & -ij_1 & & \cdots &&&&&\\
& ij_1 & r_b & -ij_1 & \cdots &&&&& \\
&&&& \ddots &&&&\\
&&&& \cdots & ij_1& r_b & -ij_1 & r_{0,2} & \\
&&&& \cdots & & ij_1 & r_1 & r_{0,1}-ij_1 & \\
&&&& \cdots & & r_{0,2} & r_{0,1}+ij_1 & r_0 & \\
&&&&\cdots&&&&& r_s
\end{pmatrix}
\end{equation}
The lattice populations have four possible values, corresponding to boundary ($r_{K-1},r_1,r_0$) and bulk ($r_b$) values, given by
\begin{align} \label{lattice_populations}
\begin{split}
&r_{K-1}=\frac{1}{2}\left(\frac{1}{2\delta}\right)^2,\qquad r_b = (1+\Gamma^2)r_{K-1},\\
&r_1=\Gamma^2r_{K-1},\qquad r_0=-r_{K-1}(1+\Gamma^2)(K-2)-\frac{1}{2}.
\end{split}
\end{align}
The off-diagonal real values are
\begin{equation}
r_{0,1}=-\frac{1}{4\delta},\qquad r_{0,2}= -\frac{1}{2}\left(\frac{1}{2\delta}\right)^2.
\end{equation}
The imaginary values correspond to the expectation value of the (homogeneous) NESS current, given by
\begin{equation}
j_1=\frac{\langle\hat{\J}_{\text{min},k}\rangle}{\mu}=\frac{1}{2}\Gamma\left(\frac{1}{2\delta}\right)^2.
\end{equation}
Finally, the population of the auxiliary state is $r_s=1/2$. We have verified numerically that these results indeed correspond to the correct solution to the linear system of equations, and thus give the correct form of the first-order correction $\hat{\rho}^{(1)}$.

\subsection{LGIs violation enhancement by driving} \label{appendix_toy_LGI}
Here we discuss the Leggett-Garg functions of the NESS of the strongly-driven minimal model. We first consider the $f=1$ limit. Applying the operator $\hat{Q}=\hat{\mathcal{I}}-2|P\rangle\langle P|$ to $\hat{\rho}^{(0)}$ we get
\begin{equation}
\hat{Q}\hat{\rho}^{(0)} = \hat{\rho}^{(0)}-\frac{2}{\mathcal{N}}\sum_{n=0}^{K-1}(2\delta)^{-(K-P+n)}|P\rangle\langle K-n|.
\end{equation}
The dominant term of the sum is of $\mathcal{O}((2\delta)^{-(K-p)})$ (for $n=0$), which is insignificant within the order $O(\delta^{-2})$ of our solution if site $P$ is far from the right boundary. Thus $\hat{Q}\hat{\rho}^{(0)}\approx\hat{\rho}^{(0)}$, and the time correlation is
\begin{align}
\mathcal{C}^{(0)}(t)=\tr\left(\hat{Q}\exp{[\hat{\mathcal{L}}^{(0)}t]}\hat{Q}\hat{\rho}^{(0)}\right)\approx\tr\left(\hat{Q}\exp{[\hat{\mathcal{L}}^{(0)}t]}\hat{\rho}^{(0)}\right)=\tr\left(\hat{Q}\hat{\rho}^{(0)}\right)\approx\tr(\hat{\rho}^{(0)})=1.
\end{align}
It is approximately constant, as expected from the insulating nature of the state. The corresponding Leggett-Garg function is
\begin{equation}
\mathcal{K}^{(0)}(t)=2\mathcal{C}^{(0)}(t)-\mathcal{C}^{(0)}(2t)\approx1,
\end{equation}
indicating that the LGI is not violated\footnote{We have observed numerically that $\mathcal{K}^{(0)}(t)$ is actually slightly larger than one for short times, but since the governing term is of $\mathcal{O}((2\delta)^{-(K-P)})$, such a violation is insignificant for $\delta>1$.}.\\

Now we consider the system slightly below maximal driving, namely with $\mu=1-f\ll1$, whose NESS was discussed in Section~\ref{appendix_toy_pert}. With the NESS of Eq.~\eqref{pertur_mu} calculated up to first order in $\mu$ we can proceed to obtain the perturbative correction to the early-time time correlations $\mathcal{C}^{(1)}(t)$ and Leggett-Garg function $\mathcal{K}^{(1)}(t)$, so that
\begin{align}
\mathcal{C}(t)&=\mathcal{C}^{(0)}(t)+\mu\mathcal{C}^{(1)}(t),\\
\mathcal{K}(t)&=\mathcal{K}^{(0)}(t)+\mu\mathcal{K}^{(1)}(t).
\end{align}
Expanding the time correlations in Eq. (3) of the main text up to second order in time, the perturbative correction to the $f=1$ Leggett-Garg function is
\begin{equation} \label{2dnorderK}
\mathcal{K}^{(1)}(t)=-t^2{\rm Re}\left(\tr\pas{\hat{Q}\hat{\mathcal{L}}^2\hat{Q}\hat{\rho}^{(1)}}\right), 
\end{equation} 
indicating that there is no linear correction in time for $\mathcal{K}$. To calculate it, first we note that applying $\hat{Q}$ to $\hat{\rho}^{(1)}$ gives
\begin{equation}
\hat{Q}\hat{\rho}^{(1)} = \hat{\rho}^{(1)} - 2r_b|P\rangle\langle P|+2ij_1\left(|P\rangle\langle P+1|-|P\rangle\langle P-1|\right).
\end{equation}
Applying the Lindblad operator $\hat{\mathcal{L}}$ twice to $\hat{Q}\hat{\rho}^{(1)}$ is a lengthy process, so we focused on obtaining the real diagonal terms which give a finite contribution to the trace after being multiplied again by $\hat{Q}$. We finally find that 
\begin{align}
{\rm Re}\left(\tr\pas{\hat{Q}\hat{\mathcal{L}}^2\hat{Q}\hat{\rho}^{(1)}}\right)=-2{\rm Re}\left(\tr|P\rangle\langle P|2r_b|P\rangle\langle P|\right)=-4r_b,
\end{align}
which, using Eq.~\eqref{lattice_populations}, gives that the correction of the LGI function is
\begin{equation} \label{lgi_toy_mu}
\mathcal{K}(t)-\mathcal{K}^{(0)}(t)=2(1-f)(1+\Gamma^2)\left(\frac{t}{2\delta}\right)^2.
\end{equation}
Since this is a positive quantity and $\mathcal{K}^{(0)}(t)\approx1$, we indeed observe that LGIs are violated when moving slightly away from the $f=1$ scenario, and that such violation increases with time and as $f$ decreases. In Fig.~\ref{fig_4_sm}(a) we show that this prediction agrees well with exact numerical results for early times.

\subsection{LGIs violation enhancement by dephasing} \label{appendix_toy_dephasing}
An identical calculation can be performed for the maximally-driven $f=1$ limit, but with a weak dephasing rate $\gamma\ll1$. However this can also be obtained considering the driving-dephasing equivalence in the minimal model described in Ref.~\cite{Mendoza:2013a}, both were seen to induce the same effect on the NESS when replacing  
\begin{equation}
\gamma=\frac{1}{2}(1-f)\frac{\Gamma}{4}.
\end{equation}
The correction to the Leggett-Garg function is then
\begin{equation} \label{lgi_toy_deph}
\mathcal{K}(t)-\mathcal{K}^{(0)}(t)=16\gamma\frac{(1+\Gamma^2)}{\Gamma}\left(\frac{t}{2\delta}\right)^2,
\end{equation}
which is also positive and thus indicates that dephasing induces a violation of the LGIs, which increases with $\gamma$. This result is also in agreement with exact numerical calculations, as depicted in In Fig.~\ref{fig_4_sm}(b). If both a weak backflow and dephasing are present, where both mechanisms enhance the LGI violations, the result is simply the sum of the independent contributions of Eqs.~\eqref{lgi_toy_mu} and~\eqref{lgi_toy_deph}, namely
\begin{equation} \label{lgi_toy_both}
\mathcal{K}(t)-\mathcal{K}^{(0)}(t)=2(1+\Gamma^2)\left[(1-f)+\frac{8\gamma}{\Gamma}\right]\left(\frac{t}{2\delta}\right)^2.
\end{equation}  
This additive enhancement of the two mechanisms is verified by exact numerical simulations, which as shown in Fig.~\ref{fig_4_sm}(c), present a very good agreement.

\begin{figure}[t]
\begin{center}
\includegraphics[scale=0.8]{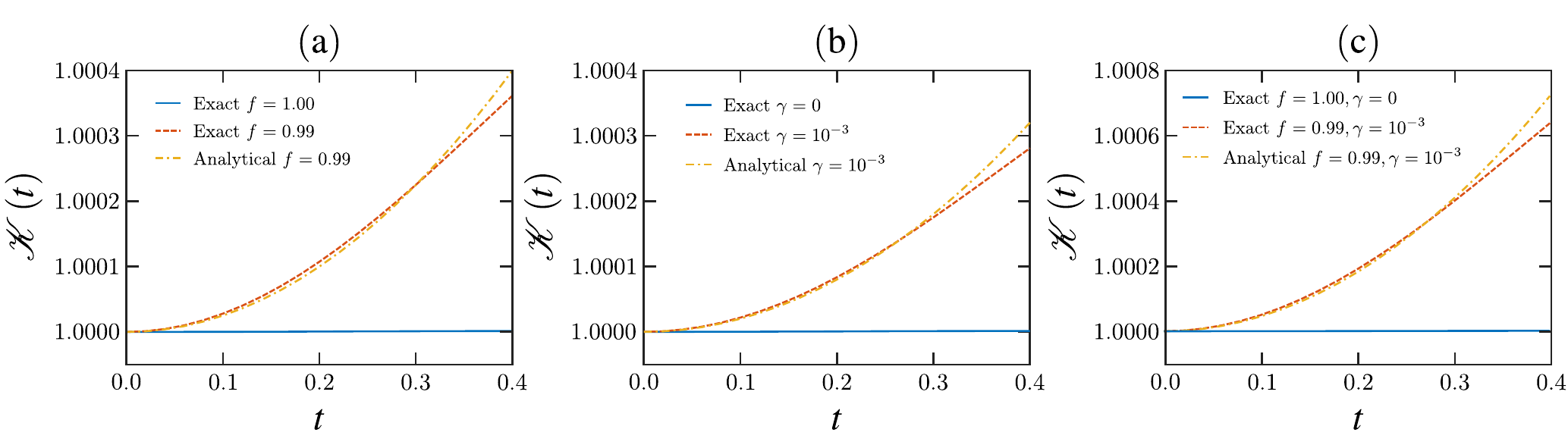}
\caption{{\bf Comparison of analytical and numerical results for the minimal model.} The results correspond to $K=10$, $\delta = 2$ and $\Gamma=1$, (a) for $f=0.99$ and $\gamma=0$, (b) for $f=1.00$ and $\gamma=10^{-3}$, (c) for $f=0.99$ and $\gamma=10^{-3}$.}\label{fig_4_sm}
\end{center}
\end{figure}

\end{document}